\documentclass{article}
\usepackage{spconfa4,amsmath,graphicx}
\usepackage{cite}
\usepackage{url}
\usepackage{makecell, hhline}
\usepackage{float}
\usepackage{amssymb,amsfonts}
\usepackage{algorithmic}
\usepackage{textcomp}
\usepackage{xcolor}
\usepackage[nolist,nohyperlinks]{acronym}
\usepackage{multirow}
\usepackage[subscriptcorrection,varg,varvw,cmbraces,cmintegrals]{newtxmath}

\title{XANE Background Acoustic Embeddings: Ablation and Clustering Analysis}

\name{\begin{tabular}{c}
     Dushyant Sharma$^{1}$, 
      James Fosburgh$^{1}$,
      Sri Harsha Dumpala$^{2}$\sthanks{Sri Harsha Dumpala and Chandramouli Shama Sastri were interns at Microsoft during the course of this work!},
      Chandramouli Shama Sastri$^{2}$,\\
     Stanislav Yu. Kruchinin$^{1}$ and Patrick A. Naylor$^{3}$
\end{tabular}}
      
\address{ $^1$Microsoft Inc.\\ 
$^2$Vector Institute, Canada\\
  $^3$Imperial College London, UK
}

\begin{document}

\ninept

\maketitle

\begin{abstract}
We explore the recently proposed explainable acoustic neural embedding~(XANE) system that models the background acoustics of a speech signal in a non-intrusive manner. The XANE embeddings are used to estimate specific parameters related to the background acoustic properties of the signal which allows the embeddings to be explainable in terms of those parameters. We perform ablation studies on the XANE system and show that estimating all acoustic parameters jointly has an overall positive effect. Furthermore, we illustrate the value of XANE embeddings by performing clustering experiments on unseen test data and show that the proposed embeddings achieve a mean F1 score of 92\% for three different tasks, outperforming significantly the WavLM based signal embeddings and are complimentary to speaker embeddings.
\end{abstract}
%
%
\section{Introduction}
\label{sec:intro}
A speech signal acquired in the real world may be adversely affected by additive environmental noise, room reverberation and CODEC artifacts. Previous studies have shown that accurate estimation of these parameters can be beneficial for ASR~\cite{nisapp,parada2015reverberant} and other speech processing tasks~\cite{kitawaki1988quality,hu2015speaker}. The effect of acoustic reverberation is typically modeled as the convolution between an anechoic speech signal and a Room Impulse Response~(RIR)~\cite{naylor2010speech}. A number of parameters characterizing the RIR have been proposed in the literature, including the Clarity index~($C_{50}$), reverberation time~($T_{60}$) and direct-to-reverberant ratio~(DRR)~\cite{parada2014non}. In~\cite{hu2015speaker}, it was shown that the $C_{5}$ metric is valuable for speaker diarization. The noise level is typically characterized by the Signal-To-Noise Ratio~(SNR) and the combination of all degrading effects can measured from a perceptual quality and intelligibly perspective by methods such as PESQ~\cite{recommendation2001perceptual} and ESTOI~\cite{estoi}. The problem of Voice Activity Detection~(VAD) is also closely linked with the accurate estimation of background acoustic parameters. Over the past decade, a number of algorithms have been proposed for the task of non-intrusive signal analysis~\cite{parada2014non, Sharma2020-NIE, Gamper2018}, including methods for estimating reverberation parameters~\cite{parada2014non, Gamper2018, Sharma2020-NIE}, objective speech quality and intelligibility~\cite{mittag2021nisqa, yi2022conferencingspeech} and the bit rate of speech CODECs~\cite{sharma2017non}. In our previous work, we proposed the eXplainable Acoustic Neural Embeddings~(XANE)~\cite{xane_interspeech} method, which extracts a neural embedding that encapsulates information about the background acoustics in a speech signal in the form of a vector representation. XANE operates in a non-intrusive framework and makes the embeddings explainable by further estimating a wide range of background acoustic parameters from the neural embeddings, using a Transformer based architecture. In this work, we analyze the XANE embeddings and compare them with WavLM~\cite{chen2022wavlm} and speaker embeddings from ~\cite{heo2020clova} using t-SNE (t-distributed Stochastic Neighbor Embedding)~\cite{tsne}, which is a technique for visualizing high-dimensional data in a lower-dimensional space and k-means~\cite{kmeans} clustering. We also explore the impact of different embedding dimension on the estimation of the 14 acoustic parameters, as well as ablation studies to evaluate the impact of removing noise type, CODEC type and overlapped speech classification objectives from the training process.

\section{XANE System}\label{sec:nisa-emb}
\begin{figure}[h]
\centering
  \centerline{\includegraphics[width=0.99\linewidth]{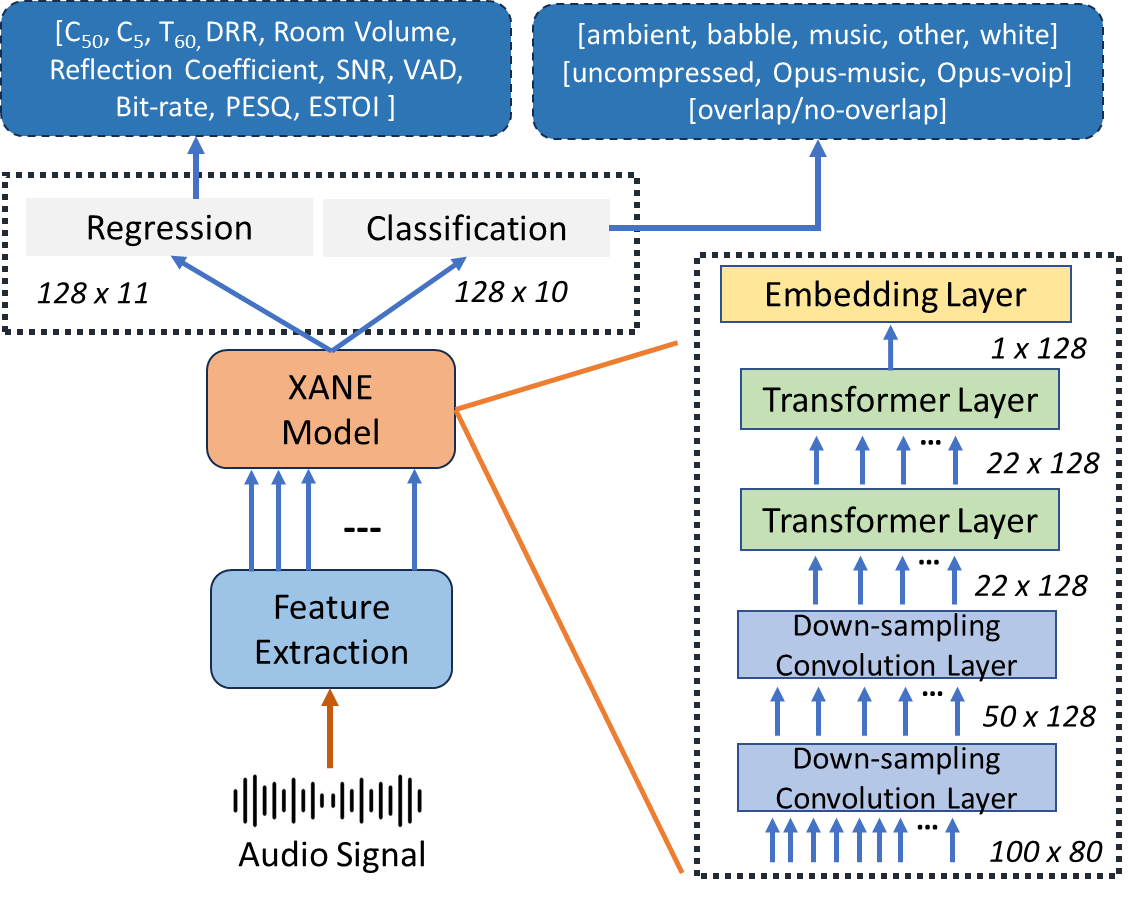}}
  \caption{XANE architecture based on Transformers and MelFB features.}
  \label{fig:nisa_outline}
\end{figure}

Figure~\ref{fig:nisa_outline} shows the outline of the XANE architecture, using Mel Filterbank~(MFB) features and a transformer based neural network as that was found to be the best system in our previous work~\cite{xane_interspeech}. All signals were sampled at 16 kHz and 80 dimension MelFB features were extracted using a frame size of 25~ms and 10~ms frame increment. XANE uses a context or chunk size of 100 frames (corresponding to 1~s of audio) to estimate 14 acoustic parameters from an embedding layer that has a dimension of 128. The Transformer model consists of downsampling convolutional blocks which consist of two convolutional~(Conv.) layers (256 channels with a stride of 2) that reduce the input frame rate by a factor of 4. The output from these is processed through a Transformer block with two encoder layers with 256-dimensional input, 8 attention heads and a 256-dimensional fully connected linear layer. The output of the Transformer block is passed through a fully connected layer with 128 units (the embedding layer) and the GELU activation function~\cite{Hendrycks2016Jun}, followed by the output layer. The output layer, performs estimation of 11 regression tasks and three classification tasks. The classification tasks are (1) noise type classification (comprising ambient, babble, music, other and white noise), (2) CODEC type classification (uncompressed, Opus ``music'' and ``speech' presets) and (3) speech overlap detection (overlapped or non-overlapped). The regression tasks include $C_{50}$, $C_{5}$, $T_{60}$, DRR, room volume and reflection coefficient estimation in addition to SNR, VAD, PESQ~\cite{recommendation2001perceptual}, ESTOI~\cite{estoi} and CODEC bit rate. 
\section{Data and Evaluation}\label{data}
The XANE system is trained in a supervised manner, using clean speech from the training partitions of the VCTK~\cite{yamagishi2019vctk} and Timit~\cite{TIMIT} datasets, that are subsequently corrupted by various data augmentations. We convolved the clean speech with simulated RIRs covering a large configuration of room volumes, reflection coefficients and source-microphone positions, using the image method~\cite{Allen1979}. For the overlap conditions, we added speech from a different speaker to the target speaker in an utterance~(3-12~dB range). We then added ambient, babble, white, music and other (mainly domestic noises) noise in 0-30~dB SNR range and processed the audio through the respective CODEC and level augmentation in the range of -0.1 to -10~dBFS. The training dataset was organized into 6 groups, consisting of the three CODEC conditions (uncompressed, Opus ``music'' and Opus ``speech'', 8 to 64~kbps) and the two overlap conditions (overlapped or non-overlapped speech). For each of these groups, we sampled 40k utterances from the clean set and performed the RIR, noise and CODEC augmentations mentioned, resulting in a final set with 240k utterances (223.8~hrs). Clean speech from the test partitions of the VCTK and Timit datasets was used as base material for synthesising the VCTK test data for the proposed methods and followed the same pattern as the training data, but care was taken to ensure no overlap in speech material or noise and RIR  between the two sets. In addition, we used the ACE~\cite{Eaton2016} dataset for measuring generalization performance of the methods on some of the reverberation metrics as this dataset contains measured RIRs. We use the Mean Absolute Error~(MAE) metric for the regression tasks and F1 score for classification and clustering tasks similar to~\cite{nisapp}. 
\begin{table}[h]
\caption{Acoustic parameter estimation performance for different XANE embedding dimensions, from 32 to 512.}
\label{table_emb_dim}
\begin{tabular}{|l|c|c|c|c|c|}
\cline{2-6} \cline{3-6} \cline{4-6} \cline{5-6} \cline{6-6} 
\multicolumn{1}{l|}{} & \multicolumn{5}{c|}{\textbf{XANE Embedding Dimension}}\tabularnewline
\cline{2-6} \cline{3-6} \cline{4-6} \cline{5-6} \cline{6-6} 
\multicolumn{1}{l|}{} & \textbf{32} & \textbf{64} & \textbf{128} & \textbf{256} & \textbf{512}\tabularnewline
\hline 
\textbf{C50 (dB)} & 3.8 & 3.3 & 3.2 & \textbf{3.1} & 9.8\tabularnewline
\hline 
\textbf{T60 (ms)} & 135 & 109 & 112 & \textbf{84} & 152\tabularnewline
\hline 
\textbf{DRR (dB)} & 2.3 & \textbf{2.1} & \textbf{2.1} & 2.5 & 5.2\tabularnewline
\hline 
\textbf{C5 (dB)} & 2.2 & 2.0 & \textbf{1.9} & \textbf{1.9} & 4.5\tabularnewline
\hline 
\textbf{Rvol. ($\text{m}^{3}$)} & 5.0 & 5.0 & \textbf{4.9} & 5.6 & 8.5\tabularnewline
\hline 
\textbf{Refc. ($10^{-3}$)} & 71 & 58 & \textbf{56} & 58 & 2.25\tabularnewline
\hline 
\textbf{PESQ} & 0.33 & 0.40 & \textbf{0.31} & 0.49 & 1.12\tabularnewline
\hline 
\textbf{ESTOI ($10^{-3}$) } & 88 & 78 & 75 & \textbf{73} & 277\tabularnewline
\hline 
\textbf{BR (kbs)} & 10.5 & 11.1 & \textbf{10.3} & 10.4 & 147.1\tabularnewline
\hline 
\textbf{SNR (dB)} & 4.3 & 3.7 & \textbf{3.5} & 3.9 & 5.1\tabularnewline
\hline 
\textbf{Noise Type } & 61.6 & 62.6 & \textbf{66.4} & 59.5 & 44.2\tabularnewline
\hline 
\textbf{CODEC Type} & 99.3 & \textbf{99.8} & 99.7 & 99.4 & 67.7\tabularnewline
\hline 
\textbf{Overlap Det.} & 90.0 & 91.4 & \textbf{92.4} & 92.3 & 50.0\tabularnewline
\hline 
\multicolumn{1}{|c|}{\textbf{\# Param. (M)}} & 0.57 & 0.67 & 0.97 & 3.10 & 14.65\tabularnewline
\hline 
\end{tabular}
\caption{Clustering experiments (F1 score). The XANE-(NN/NC/NO) conditions represent models trained without noise, CODEC or overlap classification, respectively.}
\label{tab:kmeans_clustering}
\begin{tabular}{|c|c|c|c|c|c|}
\cline{2-6} \cline{3-6} \cline{4-6} \cline{5-6} \cline{6-6} 
\multicolumn{1}{c|}{} & \multicolumn{2}{c|}{\textbf{ACE}} & \multicolumn{3}{c|}{\textbf{VCTK}}\tabularnewline
\cline{2-6} \cline{3-6} \cline{4-6} \cline{5-6} \cline{6-6} 
\multicolumn{1}{c|}{} & \textbf{Noise} & \textbf{Reverb} & \textbf{Noise} & \textbf{Reverb} & \textbf{Overlap}\tabularnewline
\hline 
\textbf{WavLM} & 0.68 & 0.55 & 0.38 & 0.81 & 0.51\tabularnewline
\hline 
\textbf{XANE} & 0.86 & 0.99 & 0.77 & 1.00 & 0.98\tabularnewline
\hline 
\textbf{XANE-NN} & 0.65 & 0.99 & 0.48 & 0.91 & 0.97\tabularnewline
\hline 
\textbf{XANE-NC} & 0.86 & 0.99 & 0.73 & 1.00 & 0.97\tabularnewline
\hline 
\textbf{XANE-NO} & 0.85 & 0.99 & 0.69 & 1.00 & 0.62\tabularnewline
\hline 
\textbf{Spk} & 0.42 & 0.50 & 0.22 & 0.51 & 0.52\tabularnewline
\hline 
\end{tabular}
\vspace{-15pt}
\end{table} 
\section{Experiments and Results}\label{experiments}
\subsection{Embedding Dimension}
In the first set of experiments, we evaluated the impact of different embedding dimensions on the estimation of the 14 acoustic parameters. As shown in Table~\ref{table_emb_dim}, we evaluated embedding dimensions from 32 to 512. In order to support the different embedding dimensions, we also modified the downsampling Conv. layers and the feed-forward layer dimension in the transformer architecture appropriately. This leads to the model with a 32 dimension model to only have 0.57~Million learnable parameters, compared with 14.65~Million for the 512 dimension model. We can see that the XANE model with 128 dimension embeddings achieves the best overall accuracy, outperforming the other models in the DRR, $C_{5}$, room volume, reflection coefficient, PESQ, bit rate, SNR, noise type classification and overlapped speech detection tasks.

\subsection{Ablation Studies}
We also performed three ablation studies to evaluate the impact of removing the three classification tasks on the overall performance of the system. For these, we used the XANE system with the 128 dimension embeddings. In Table~\ref{table_ablation} we present the parameter estimation results for these ablation studies. We can see that the best performance overall is achieved when these tasks are included in the model training objective. Similarly, in Table~\ref{tab:kmeans_clustering} we can observe the impact on clustering of excluding the classification tasks, where again, no gain is observed in removing the tasks.
\begin{figure*}[h]
  \centering
  \begin{tabular}{ccc}
    \includegraphics[width=0.3\textwidth]{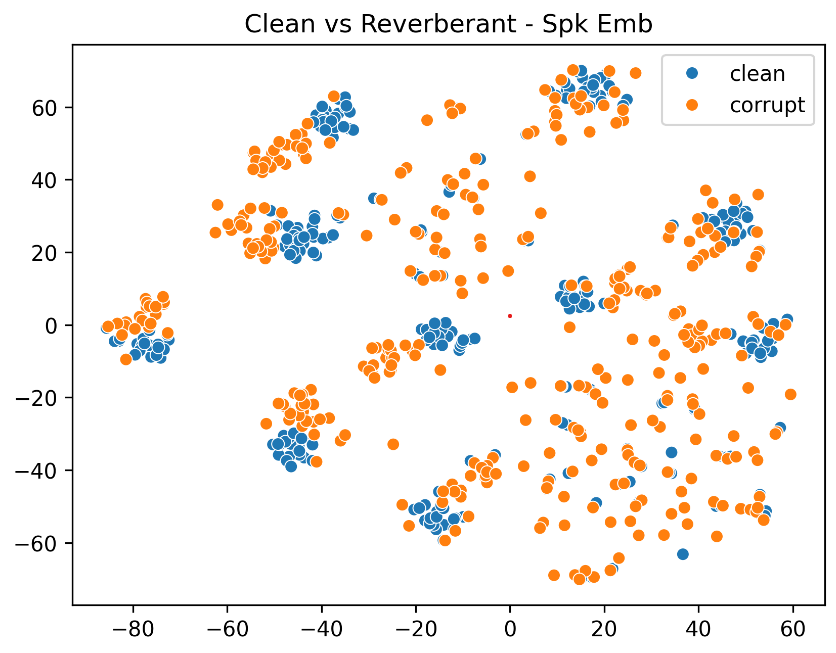} &
    \includegraphics[width=0.3\textwidth]{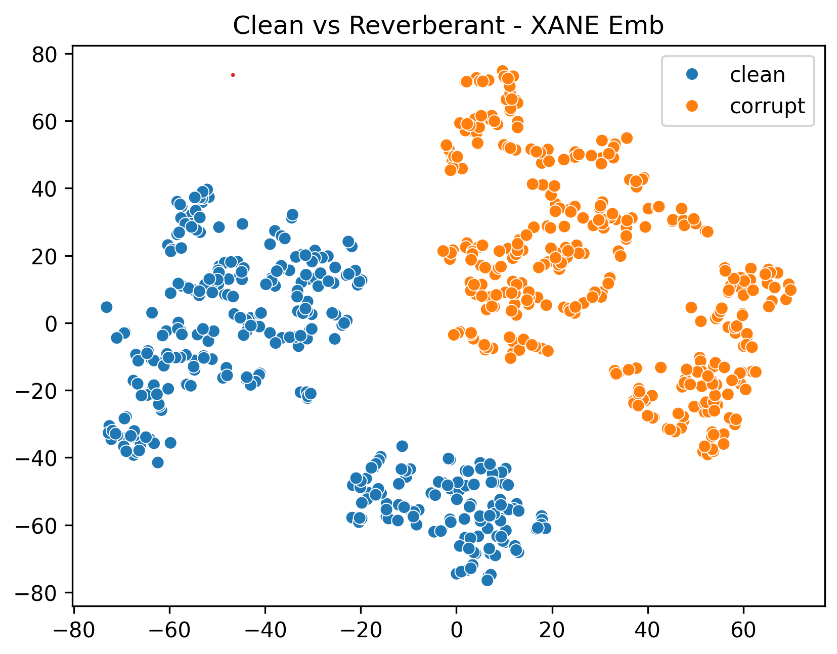} &
    \includegraphics[width=0.3\textwidth]{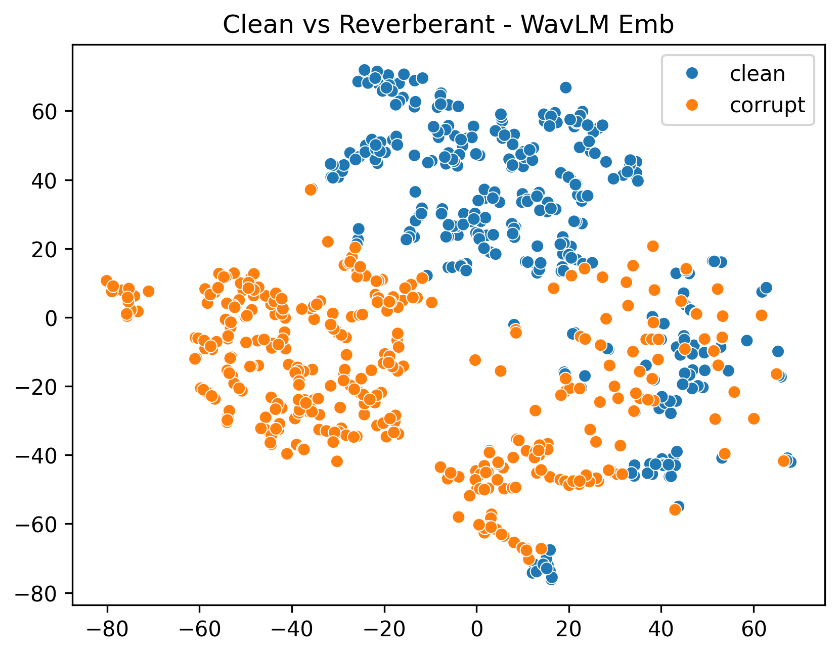} \\
    \includegraphics[width=0.3\textwidth]{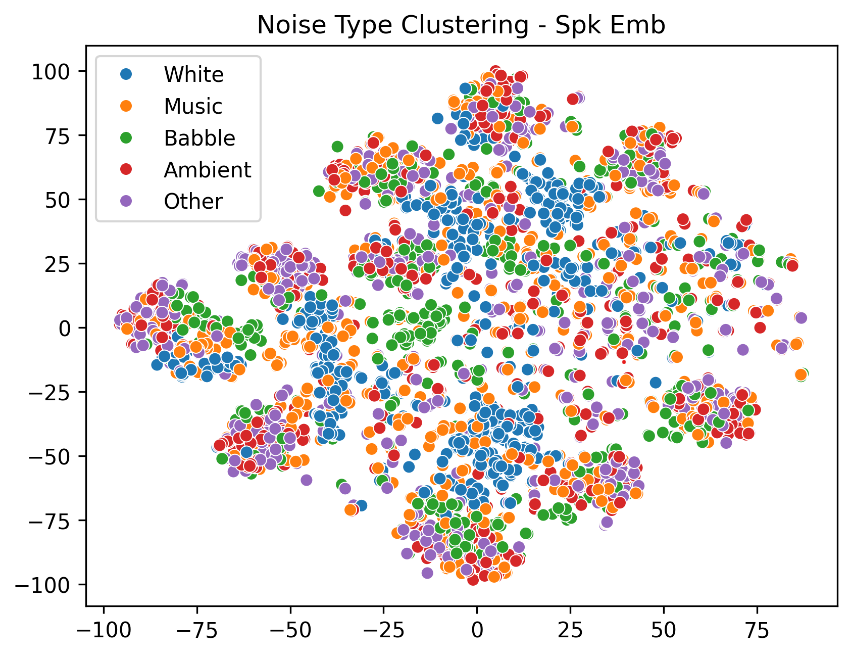} &
    \includegraphics[width=0.3\textwidth]{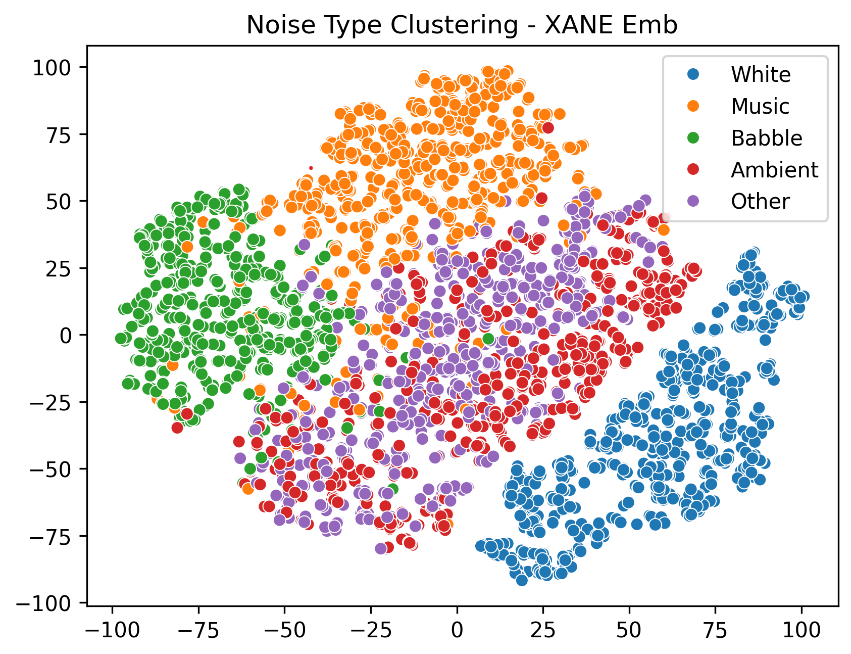} &
    \includegraphics[width=0.3\textwidth]{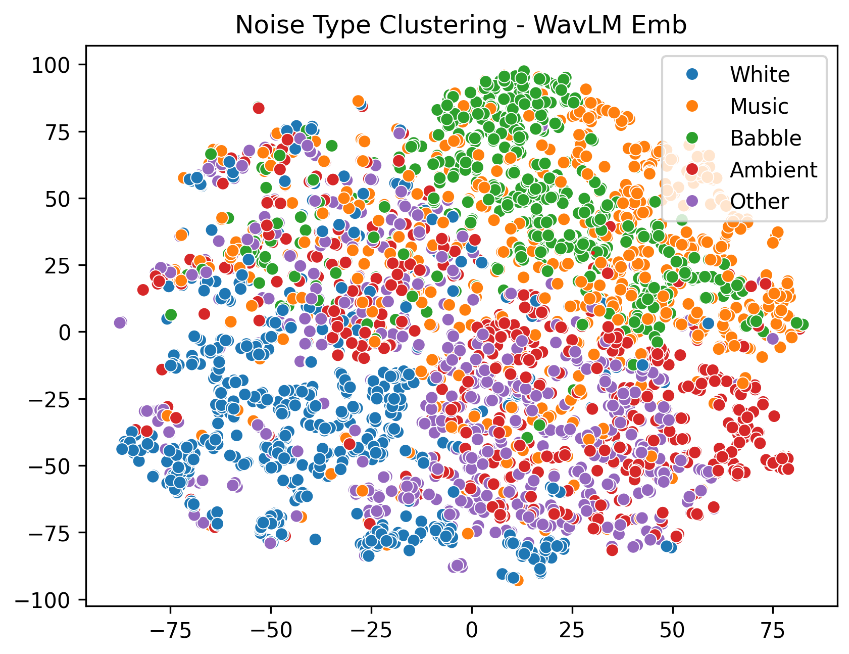}\\
  \end{tabular}
  \caption{T-SNE~\cite{tsne} plots for XANE embeddings and speaker embeddings on the VCTK based test set. The three columns represent (from left to right), anaechoic speech detection, noise type classification and overlapped speech detection, respectively. The top row represents the speaker embeddings and the bottom row the XANE embeddings.}
  \label{fig:tsne}
\end{figure*}
\subsection{Clustering XANE Embeddings}
To evaluate the effectiveness of the XANE embeddings, we used the T-SNE~\cite{tsne} dimensionality reduction algorithm to visualize the embeddings in a 2-dimension plot. We also evaluate two additional embedding systems: (1) a speaker embedding system based on the ResNet speaker encoder architecture as implemented in~\cite{heo2020clova}, and (2) WavLM~\cite{chen2022wavlm} based acoustic embeddings as baseline methods for comparison. We note that the speaker embeddings are not generally designed to model the background acoustics of a speech signal and we include them here as a means to highlight the complimentary nature of the acoustic embeddings. We evaluated clustering performance for noise type, presence of reverberation, and presence of overlapped speech on the VCTK test set, and noise type and presence of reverberation on the ACE test set. In each case, we sub-select from the test set to attempt to control for the acoustic property being evaluated. When evaluating for reverberation, we selected only corrupted utterances with SNRs above 20dB of a single type of noise. For the VCTK based test set, we additionally select only uncompressed utterances without overlapped speech. When testing for noise clustering, we consider corrupt utterances with an SNR below 20dB. On the VCTK test set we again consider only uncompressed utterances without overlap, and for the ACE test set we consider only uttreances with the ``highq-opus-voip" CODEC. Finally, when evaluating overlap speech clustering on the VCTK test set, we only evaluate uncompressed utterances with white noise. As can be seen in Fig.~\ref{fig:tsne}, the visualizations of the XANE embeddings show distinct clusters when viewing plots for presence of reverberation and overlap speech, while the speaker embeddings do not, and have less overlap in clusters than those for WavLM. In noise classification, the most significantly overlapped groups for the XANE embeddings are ``Ambient" and ``Other". In this test set, ``other" is comprised of fan noise, which bares a very strong acoustic resemblance to ambient noise. We additionally present the F1 score of clusters created with a k-means~\cite{kmeans} algorithm, as shown in Table~\ref{tab:kmeans_clustering}. Here we additionally evaluated the models from the ablation study. As can be seen, the XANE embeddings outperform the the speaker embeddings as expected, and additionally outperform the WavLM embeddings in every case. We note that, in the case of clustering by reverberation on the VCTK based test set, the performance of the XANE-NN embeddings is lower than the other XANE embeddings. This likely means that the other XANE embeddings are still picking up on the high SNR white noise present in the test utterances, and thus the XANE-NN performance is likely most representative of truly clustering only on presence of reverberation. Finally, as the XANE system is trained to be speaker agnostic, we expect the embeddings for utterances from different speakers with the same acoustic environments to be very similar. To evaluate this, we pick one utterance from each speaker in the clean version of the VCTK based test set. Choosing one speaker as the reference, we then compute the cosine distance between the embedding from the reference and the embedding from each other speaker. As can be seen in Table~\ref{table:speaker_distance}, the XANE embeddings have the lowest mean cosine distance between speakers.


\begin{table}
\centering
\vspace{-5pt}
\caption{Mean and std. in cosine distance between embeddings across speakers in the VCTK based test set.}
\begin{tabular}{|c|c|c|c|}
\hline
& \textbf{XANE} & \textbf{WavLM} & \textbf{Spk} \tabularnewline
\hline
\textbf{Mean} & 0.16 & 0.19 & 0.85 \tabularnewline
\hline
\textbf{Std.} & 0.06 & 0.04 & 0.09 \tabularnewline
\hline
\end{tabular}
\label{table:speaker_distance}
\vspace{-5pt}
\end{table}

\begin{figure}[h]
  \centering
   \includegraphics[width=0.35\textwidth]{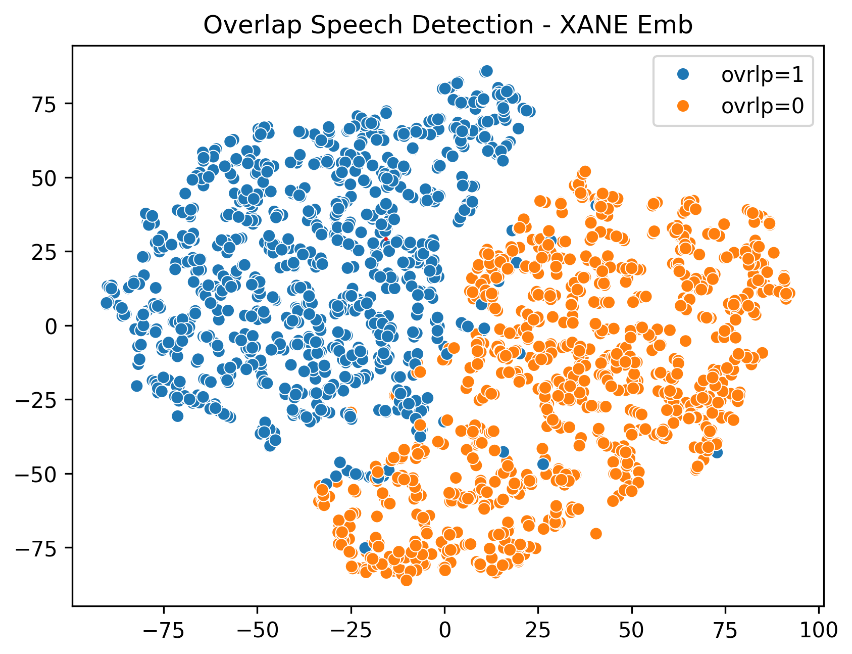}\\
  \caption{T-SNE~\cite{tsne} plots for XANE and WavLM embeddings on the VCTK based test set for the overlapped speech condition. }
  \label{fig:tsne_overlap}
\end{figure}

\begin{table}[h]
\vspace{-5pt}
\caption{Acoustic parameter estimation performance for different ablation conditions (no noise, no overlapped speech and no CODEC classification), using 128 dimension XANE embeddings.}
\label{table_ablation}
\begin{tabular}{|l|c|c|c|c|}
\cline{3-5} \cline{4-5} \cline{5-5} 
\multicolumn{1}{l}{} & \multicolumn{1}{c|}{} & \multicolumn{3}{c|}{\textbf{Ablation Condition}}\tabularnewline
\cline{2-5} \cline{3-5} \cline{4-5} \cline{5-5} 
\multicolumn{1}{l|}{} & \textbf{Base} & \textbf{--Noise} & \textbf{--Overlap} & \textbf{--CODEC}\tabularnewline
\hline 
\textbf{C50 (dB)} & 3.2 & \textbf{2.8} & 3.4 & 2.9\tabularnewline
\hline 
\textbf{T60 (ms)} & 112 & \textbf{93} & 100 & 102\tabularnewline
\hline 
\textbf{DRR (dB)} & 2.1 & 2.1 & 2.1 & \textbf{2.0}\tabularnewline
\hline 
\textbf{C5 (dB)} & \textbf{1.9} & \textbf{1.9} & \textbf{1.9} & \textbf{1.9}\tabularnewline
\hline 
\textbf{Rvol. ($\text{m}^{3}$)} & \textbf{4.9} & \textbf{4.9} & \textbf{4.9} & 5.2\tabularnewline
\hline 
\textbf{Refc. ($10^{-3}$)} & 56 & 63 & 58 & \textbf{48}\tabularnewline
\hline 
\textbf{PESQ} & 0.31 & \textbf{0.29} & 0.33 & 0.35\tabularnewline
\hline 
\textbf{ESTOI ($10^{-3}$)} & 75 & \textbf{72} & 73 & 76\tabularnewline
\hline 
\textbf{BR (kbs)} & \textbf{10.3} & 10.8 & 10.6 & 10.5\tabularnewline
\hline 
\textbf{SNR (dB)} & \textbf{3.5} & 4.1 & 4.1 & 3.6\tabularnewline
\hline 
\textbf{Noise Type} & \textbf{66.4} & 18.2 & 58.7 & 61.6\tabularnewline
\hline 
\textbf{CODEC Type} & \textbf{99.7} & \textbf{99.7} & 98.6 & 32.8\tabularnewline
\hline 
\textbf{Overlap Det.} & \textbf{92.4} & 91.5 & 50.4 & 92.3\tabularnewline
\hline 
\end{tabular}
\end{table}
\section{Conclusions}\label{conclusions}
We presented clustering analysis and ablation studies for the XANE method that estimates neural embeddings characterizing the background acoustic from a speech signal in a non-intrusive manner and making them explainable by estimating a large set of acoustic parameters. It was found that joint estimation of the acoustic parameters results in the best performance for the system. We also compared the clustering of XANE embeddings with the WavLM and speaker embeddings and showed that XANE embeddings have a better clustering performance according to various acoustic conditions and moreover, are invariant to speaker changes. This highlights the complimentary nature of XANE embeddings when compared with speaker embeddings, and can be leveraged in future work on improved methods for text to speech synthesis, for example.

\bibliographystyle{IEEEbib}
\bibliography{refs}

\end{document}